\documentclass[12pt, a4paper,oneside]{amsart}
\usepackage{amscd}
\usepackage[mathscr]{eucal}
\def\Tr{\operatorname{\mathrm {Tr}}\nolimits}
\begin{document}
\newtheorem{theorem}{Theorem}[section]
\newtheorem{lemma}{Lemma}[section]
\newtheorem{definition}{Definition}[section]
\newtheorem{corollary}{Corollary}[section]
\newtheorem{proposition}{Proposition}[section]

\title[Quantum Communication and  Multivariate Interpolation]{Quantum Communication and Quantum Multivariate Polynomial Interpolation}        
\author{Do Ngoc Diep${}^{1,3}$, Do Hoang Giang${}^2$} 
\date{\today}
\address{${}^1$ Institute of Mathematics, Vietnam Academy of Sciences and Technology, 18 Hoang Quoc Viet road, Cau Giay District, 10307 Hanoi, Vietnam}
\email{dndiep@math.ac.vn}
\address{${}^2$ K47A1T, Department of Mathematics, Mechanics and Informatics, College of Natural Sciences, Vietnam National University, 40 Nguyen Trai road, Thanh Xuan District, Hanoi Vietnam.}
\email{dhgiang84@gmail.com}
\address{${}^3$ Thang Long University, Nghiem Xuan Yem road, Hoang Mai district, Hanoi, Vietnam}
\maketitle
\begin{abstract}  
The paper is devoted to the problem of multivariate polynomial interpolation and its application to quantum secret sharing. We show that using quantum Fourier transform one can produce the protocol for quantum secret sharing distribution. 
\end{abstract}

\subjclass{\sl AMS Mathematics Subject Classification:} 15A06; 15A99

\keywords{\sl Keywords and Terms:} quantum algorithm; quantum secret sharing scheme; quantum multivariate interpolation

\section{Introduction}     

The following problem is normally considered in classical and quantum communication scenarios: Bob have to communicate some secret message to Eve. They proceed the following procedure. 
\begin{itemize}
\item \textit{Encoding message and sending}: Eve prepares some text message, namely in state $|\psi_0\rangle$, encodes it into a state $|\psi_1\rangle$ and send it to Bob
\item \textit{Queries and Back Sending }:
Bob makes some queries and transforms the received message into the state $|\psi_2\rangle$ and sends it back to Eve.
\item \textit{Decoding and Identifying the Secret}:
Eve decodes it and identifies the common secret share with Bob.
\end{itemize}

In their paper \cite{naganaka} K. Nagata and T. Nakamura described this procedure of quantum communication by using a scheme with Hadamard inverse transform $H^{\otimes N}$, N queries Q: $(x_i,y_i), i=\overline{1,N}$ and Hadamard gates $H^{\otimes N}$ :
$$\CD |\psi_0\rangle @>H^{\otimes N}>>  |\psi_1\rangle @>Q: (x_i,y_i),i=\overline{1,N}>> |\psi_2\rangle @>H^{\otimes N}>> |\psi_3\rangle \endCD$$
This scheme is applicable to the Bernstein-Vazirani algorithm for linear black boxes $f$ for finding their coefficients as the secret. 

In this paper we consider the Quatum Fourier and the inverse Quantum Fourier Trasnforms $QFT$, $QFT^{-1}$  in place of the Hadamard and inverse Hadamard gates and we discover that it is applicable to nonlinear black box function $f$ of multivariate interpolation of polynomials in higer degrees. Our scheme works for nonlinear qudits in the same way as in the linear case for qubits because for qubits, $d=2$ we have $(-1)^{xz} = \exp(i2\pi xz/2)$ and the Quantum Fourier transform
$|x\rangle \mapsto \sum_{z\in \mathbb F_q^k} e(xz),$ with $e(z) := \exp(i2\pi \Tr(z)/p)$, $\Tr(z) := z + z^{p^1} + \dots + z^{p^{d-1}}$ 
 for qubits (d=2) becomes the Hadamard gate. 

In the work \cite{chenetc}  J. Chen, A. M. Childs and S.-H. Hung  estimated the queries complexity of order $k_{\mathbb F_{q=p^r}} = \frac{d}{n+d}\binom{n+d}{d}$ with probability $1-O(1/q)$ and this complexity is \textit{optimal}. Therefore we may use it to make distribution of secret scheme among $k=k_{\mathbb F_q}$ users.

A novelty in the paper is that we use the multivariate polynomial interpolation in place of linear interpolation, qudits in place of qubits and apply to quantum key distribution for general interpolation as in the scheme of Shamir's code or NSA codes.

The paper is organized as follows. In the next section 2 we describe the quantum communication scheme, needed for our problem. In section 3 we describe the quantum multivariate polynomial interpolation and then use it to the problem of quantum secret sharing communication.

\section{Quantum communication}

The black box $U_f$ is given by a function $$f: \mathbb Z_2^N = \{0,1\}^N \to \mathbb Z_2=\{0,1\}; f(x) = a.x =\sum_{i=1}^N a_ix_i (\mod 2),$$ with $x, a \in \mathbb Z_2^N$.
The problem is to identify the coefficients $a_1,\dots,a_N \in \mathbb Z_2$. To solve this problem, there is well-known\\  \textbf{Bernstein-Vazirani algorithm}:
\begin{itemize}
\item \textit{Input} the state $|\psi_0\rangle = |0\dots 01\rangle \in \mathbb Z_2^{\otimes (N+1)}$. Let $H=\frac{1}{\sqrt{2}}\left[\begin{array}{cc} 1 & 1\\ 1 & -1 \end{array}\right]$ be the Hadamard gate, acting on pairs consisting of states, each from the first N qubits and the last (N+1)th qubit, $$H= H^{-1}: \left\{ \begin{array}{lll}|0\rangle & \mapsto& \frac{1}{\sqrt{2}}(|0\rangle + |1\rangle)\\    
|1\rangle & \mapsto& \frac{1}{\sqrt{2}}(|0\rangle - |1\rangle)\end{array}\right.$$ 
Apply $H^{\otimes N}$ to $|\psi _0\rangle$ to obtain the state $|\psi_1\rangle = H^{\otimes N}|\psi_0\rangle$;
$$|\psi_1 \rangle = \sum_{x\in \mathbb Z_2^N} \frac{|x\rangle}{\sqrt{2^N}} \left[ \frac{|0\rangle -|1\rangle}{\sqrt{2}} \right]$$
\item Make queries $x_i, f(x_i), i=\overline{1,N}$,  $U_f : |x,y\rangle \mapsto |x,y\oplus f(x)\rangle$ to obtain the state $|\psi_2\rangle = (H)^{\otimes N} |\psi_1\rangle$;
$$|\psi_2\rangle = \pm \sum_{x\in \mathbb Z_2^N} \frac{(-1)^{f(x)}|x\rangle}{\sqrt{2^N}}\left[ \frac{|0\rangle -|1\rangle}{\sqrt{2}} \right]$$
\item Apply the Hadamard gates $H^{\otimes N}$ to qubits to obtain the state $|\psi_3\rangle= H^{\otimes N}|\psi_2\rangle$; 
$$|\psi_3\rangle = \pm \sum_{z}\sum_{x} \frac{(-1)^{xz+f(x)} |z\rangle}{2^N}  \left[ \frac{|0\rangle -|1\rangle}{\sqrt{2}} \right]$$
$$= \pm \sum_{z}\sum_{x} \frac{(-1)^{x.z+a.x} |z\rangle}{2^N}  \left[ \frac{|0\rangle -|1\rangle}{\sqrt{2}} \right]$$
$$= \delta_{z+a, 0} |z\rangle  \left[ \frac{|0\rangle -|1\rangle}{\sqrt{2}} \right]$$
$$= \pm |a_1\dots a_N\rangle  \left[ \frac{|0\rangle -|1\rangle}{\sqrt{2}} \right].$$
Make a measurement and obtain the values $a_1,\dots,a_N \in \mathbb Z_2$.
\end{itemize}. We refer the readers to Ref \cite{naganaka} for more details.

This procedure gives us some application in secret sharing problem in quantum communication as follows. The scenario is that Bob wants to send a secret message to Eve, represented in form of coefficients of a linear function $f$. 

\textbf{The protocol} is to proceed the following.
\begin{itemize}
\item \textit{Encoding message and sending}: Eve prepares some text message, namely in state $|\psi_0\rangle = |0\dots 01\rangle \in \mathbb Z_2^{\otimes (N+1)}$, encodes it into a state $|\psi_1 = H^{\otimes N}|\psi_0\rangle$; 
$$|\psi_1 \rangle = \sum_{x\in \mathbb Z_2^N} \frac{|x\rangle}{\sqrt{2^N}} \left[ \frac{|0\rangle -|1\rangle}{\sqrt{2}} \right]$$
and send it to Bob
\item \textit{Queries and Back Sending }:
Bob makes some queries $Q: y_i = f(x_i), i=\overline{1,\dots N}$ and transforms it into the state $|\psi_2\rangle=Q|\psi_1\rangle$;  
$$|\psi_2\rangle = \pm \sum_{x\in \mathbb Z_2^N} \frac{(-1)^{f(x)}|x\rangle}{\sqrt{2^N}}\left[ \frac{|0\rangle -|1\rangle}{\sqrt{2}} \right]$$
and sends it back to Eve.
\item \textit{Decoding and Identifying the Secret}:
Eve decodes it by $|\psi_3\rangle = H^{\otimes N}|\psi_2\rangle$  and identifies the common secret share with Bob.
$$|\psi_3\rangle = \pm \sum_{z}\sum_{x} \frac{(-1)^{xz+f(x)} |z\rangle}{2^N}  \left[ \frac{|0\rangle -|1\rangle}{\sqrt{2}} \right]$$
$$= \pm |a_1\dots a_N\rangle  \left[ \frac{|0\rangle -|1\rangle}{\sqrt{2}} \right].$$
\end{itemize}
This code works well in the case of linear black box $f$.
Next we want to do the same for $f$ to be a polynomial in $n$ variable and of degree $d$.

\section{Quantum Secret Sharing Problem}

Let  $\mathbb K= \mathbb F_q$ be a finite fields of $q=p^r$ elements, $p$ being some fixed prime, and as above, $e(z) = \exp(i2\pi \Tr(z)/p)$, $\Tr(z) := z + z^{p^1} + \dots + z^{p^{d-1}}$. 
The Quantum Fourier Transform (QFT) is defined as 
$$QFT : |x\rangle \mapsto \frac{1}{\sqrt{q}}\sum_{y\in \mathbb F^k_q} e(-x.y)|y\rangle, \forall x, y\in \mathbb F^k_1.$$ 
We will produce the following operations
$$\CD |\psi_0\rangle @>QFT)>>  |\psi_1\rangle @>Q: (x_i,y_i),i=\overline{1,N}>> |\psi_2\rangle @>QFT^{-1}>> |\psi_3\rangle \endCD$$ The result is
$$\CD |x,y\rangle @>QFT>> \frac{1}{\sqrt{q}}\sum_{y\in \mathbb F^k_q} e(-x.y)|x,z\rangle 
@>Q>> \frac{1}{\sqrt{q}}\sum_{y\in \mathbb F^k_q} e(-x.y)|x,z+f(x)\rangle\endCD$$
$$\CD @>QFT^{-1}>>\frac{1}{q}\sum_{w\in \mathbb F^k_q}\sum_{y\in \mathbb F^k_q} e(-x.z)e(w(z+f(x))|x,w\rangle @= e(yf(x))|x,y\rangle.
\endCD$$ 
Producing the $k$-queries in parallel on have
$$y_if(x_i) = \sum_{i=1}^k \sum_{j\in J} y_ix_i^jc_j$$ and denote the map $$Z: \mathbb F_q^{nl} \times \mathbb F_q^k \to \mathbb F_q^J$$ the map with components $$Z(x,y)_j = \sum_{i=1}^k y_ix_i^j$$ one has $\sum_iy_if(x_i) = Z(x,y).c$ with $c\in \mathbb F_q^k$ as the coefficients vector of the polynomial $f$.
We refer the reader to the work \cite{chenetc} of J. Chen, A. M. Childs and S.-H. Hung for more details.

It is natural now to produce the same protocol for the secret sharing communication using this Quantum Multivariate Interpolation

\begin{theorem}
Bob can send a secret message to Eve and Eve can decode the secret message with probability $1-O(1/q)$.
\end{theorem}
\textit{Proof.} The problem is solved by using the following

\textbf{Procedure.} Bob and Eve agree to use polynomials of degree $d$.
\begin{itemize}
\item \textit{Encoding message and sending}: Eve prepares some text message, devides it into a vector with $k$ components, and looks at for a state,  namely in state $|\psi_0\rangle=|x,y\rangle\in \mathbb F_q^{2k}$, encodes it into a state $|\psi_1\rangle = QFT|\psi_0\rangle$; and send it to Bob.
\item \textit{Queries and Back Sending }:
Bob makes some queries $Q: y_i = f(x_i), i=\overline{1,\dots, N}$ and transforms it into the state $|\psi_2\rangle= Q|\psi_1\rangle$;  
$$|\psi_2\rangle  =\frac{1}{\sqrt{q}}\sum_{y\in \mathbb F^k_q} e(-x.y)|x,z+f(x)\rangle$$
and sends it back to Eve.
\item \textit{Decoding and Identifying the Secret}:
Eve decodes the received message and identifies the common secret share with Bob.
$$|\psi_3\rangle = \frac{1}{q}\sum_{w\in \mathbb F^k_q}\sum_{y\in \mathbb F^k_q} e(-x.z)e(w(z+f(x))|x,w\rangle = e(y.f(x))|x,y\rangle.$$
$$y_if(x_i) = \sum_{i=1}^k \sum_{j\in J} y_ix_i^jc_j$$ and  the map $$Z: \mathbb F_q^{nl} \times \mathbb F_q^k \to \mathbb F_q^J$$
$$(x,y) \mapsto Z(x,y),$$
with components $$Z(x,y)_j = \sum_{i=1}^k y_ix_i^j$$  has $\sum_iy_if(x_i) = Z(x,y).c$ with $c\in \mathbb F_q^k$ as the coefficients vector of the polynomial $f$.
\end{itemize}
The theorem is therefore proved.  \hfill$\Box$

There are the evident cases:

\textit{Example 1.} $d=1$, $n$ is arbitrary. In this case one need only one query $k= k_{\mathbb F_q} = \frac{1}{n+1}\binom{n+1}{1} = 1$.

\textit{Example 2.} $n=1$, $d$ arbitrary. In this case we have interpolation on one variable and we have $k=k_{\mathbb F_q} = \frac{d}{d+1}\binom{1+d}{d} = d$.

\textbf{Quantum Secret Sharing Schemes}
In the work \cite{smith}, A. Smith considered the general access structure for quantum secret sharing system in appearance of a third part person.
Let us remind a little bit about this QSS schemes. An adversary structure $\mathcal A \subseteq 2^P$ over the set $P$ of all players is a set of subsets of $P$ which is downward-closed under inclusion. It is $\mathcal Q^2$ if there is  no pair of two sets in $\mathcal A$ being complemetary to each-another. It is $\mathcal Q^{2*}$, if its dual $\mathcal A^* =\{ B \subseteq P; B^c \notin \mathcal A\}$ is $\mathcal Q^2$ and finally it is self-dual if it is both $\mathcal Q^2$ and $\mathcal Q^{2*}$. In the work  \cite{smith}, A. Smith had seen (Theorem 1) that \begin{quote} Given any $\mathcal Q^{2*}$ structure $\mathcal A$ on $P$ one can find a QSS scheme and if the scheme is self-dual then the scheme is a pure-state one.\end{quote}

Let $k=k_{\mathbb F_q}= \frac{d}{d+1}\binom{1+d}{d} $ be the optimal number of interpolation points as above.
\begin{theorem}
Bob can send a secret message to $k$ persons at Eve side and Eve can decode the  secret from the concatenated message from $k$ persons with probability $1-O(1/q)$.
\end{theorem}
\textit{Proof.} In the presence of some third part person,  in order to keep secret, Eve use concatenated informations from $k$ users, which provide a self-dual adversary structure $\mathcal A$ and Bob sends each interpolation information to only one of them and receive information from these users. 

Indeed, it is possible because the estimation $k=k_{\mathbb F_q}$ is optimal \cite{chenetc}, i.e. the concatenated information from any smaller number of persons can not resolve the secret (in the sense that the interpolation is impossible). Consider the adversary structure consisting of $k$ person. Bob sends each interpolation query to a unique one of them. 

The case when a third partite person receive all $k$ informations from Bob is excluded because, following the noncloning principle of quantum computing, the message is destroyed completely and it could not be arrived to Eve, the confused message appears only when at least one of information is sended to Eve and the third part person can not discover the message because he/she knows only at most $\leq k-1$ interpolation points. That means that the third partite person cannot discover the secret and the secrecy is graranteed.
The theorem is therefore proven. The probability is appeared from the multivariate polynomial interpolation.
\hfill$\Box$

\section{Conclusion}
We have seen that for any qudit message, one can use the quantum multivariate interpolation to make secrecy of sending the information following the general scheme of Quantum Secret Sharing. In case of multivariate polynomial interpolation, we have some probability of correctly decoding the message, rather than in the one variable case with certainty. 

\section*{Acknowledgments} The first author thanks VIASM for an invited scientific stay at that excellent institution.

\end{document}